# Fermi surface of three-dimensional $La_{1-x}Sr_xMnO_3$ explored by soft-X-ray ARPES: Rhombohedral lattice distortion and its effect on magnetoresistance

L.L. Lev,[1,2,*] J. Krempaský,[1] U. Staub,[1] V.A. Rogalev,[1] T. Schmitt,[1] M. Shi,[1] P. Blaha,[3] A. S. Mishchenko,[4,2] A.A. Veligzhanin,[2] Y.V. Zubavichus,[2] M.B. Tsetlin,[2] H. Volfová,[5] J. Braun,[5] J. Minár,[5,6] and V.N. Strocov[1,†]

[1] Swiss Light Source, Paul Scherrer Institute, 5232 Villigen, Switzerland
[2] National Research Centre "Kurchatov Institute", 123182 Moscow, Russia
[3] Institut für Materialchemie, Technische Universität Wien, A-1060 Wien, Austria
[4] RIKEN Center for Emergent Matter Science, 2-1 Hirosawa, Wako, Saitama 351-0198, Japan
[5] Department Chemie, Ludwig-Maximilians-Universität München, 81377 Munich, Germany
[6] New Technologies-Research Center, University of West Bohemia, 306 14 Plzen, Czech Republic

Electronic structure of the three-dimensional colossal magnetoresistive perovskite $La_{1-x}Sr_xMnO_3$ has been established using soft-X-ray ARPES with its intrinsically sharp definition of three-dimensional electron momentum. The experimental results show much weaker polaronic coupling compared to the bilayer manganites and are consistent with the GGA+$U$ band structure. The experimental Fermi surface unveils the canonical topology of alternating three-dimensional electron spheres and hole cubes, with their shadow contours manifesting the rhombohedral lattice distortion. This picture has been confirmed by one-step photoemission calculations including displacement of the apical oxygen atoms. The rhombohedral distortion is neutral to the Jahn-Teller effect and thus polaronic coupling, but affects the double-exchange electron hopping and thus the colossal magnetoresistance effect.

PACS: 79.60.-i; 75.47.Lx; 71.18.+y; 75.47.Gk;

Hole-doped manganites with the general chemical formula $(La, Sr)_xMnO_y$ (LSMO) are typical transition metal oxides (TMOs) with perovskite structure which have attracted tremendous interest due to the discovery of their colossal magnetoresistance (CMR). Coupling of charge, orbital, spin and lattice degrees of freedoms results in a rich phase diagram of these materials, extending over antiferromagnetic, ferromagnetic (FM) and paramagnetic (PM) insulating and metallic states. The electron transport in manganites is coupled to their ferromagnetism and is generally described in the framework of the double-exchange (DE) mechanism. However, an important role in physics of these materials can be played by polaronic effects coupling of the electron and lattice degrees of freedom whose driving force is the Jahn-Teller (JT) distortion. According to the theory of Millis *et al.* [1,2] supported by experiments on the bilayer $La_{2-2x}Sr_{1+2x}MnO_7$ manganites [3,4] the competition between the DE related electron itinercy and polaronic self-trapping leads to a crossover from the Fermi liquid regime in the FM metal to the polaronic regime in the poorly conducting PM state at the critical temperature $T_c$. The CMR occurs then due to the magnetic field shifting the balance of these two effects near the crossover point.

The present study focuses on the LSMO compounds with the composition $La_{1-x}Sr_xMnO_3$ which crystallize in three-dimensional (3D) cubic-like perovskite structures (3D-LSMO) with various lattice distortions including orthorhombic and rhombohedral (RH). The structural difference compared to the layered LSMO compounds immediately affects the polaronic effects. Indeed, the strength of these effects is determined by the electron-phonon interaction characterized by the dimensionless coupling parameter $\lambda_{e\text{-}ph}$ depending on the number of the nearest Mn neighbors *z*. Namely, $\lambda_{e-ph} = \sum_m f_m^2(0)/(2zMt\omega^2)$, where $f_m$ are the force functions depending on the chemical bonds of ions, $\omega$ is the phonon frequency, and $M$ the oscillator mass [5]. Hence, the 3D-LSMO compounds with their more delocalized electron system compared to the layered LSMO [3] should show significantly smaller polaronic effects because each Mn atom in a 3D compound is coupled by the DE to six Mn neighbors against four in the single-layer $La_{2-x}Sr_xMnO_4$ and five in the bilayer $La_{2-2x}Sr_{1+2x}MnO_7$.

Angle-resolved photoelectron spectroscopy (ARPES) is the most direct method to explore the electronic structure resolved in electron momentum **k**. The intrinsically 3D nature of the 3D-LSMO compounds is a complication compared to two-dimensional materials such as cuprates or the layered LSMO [3] because the conventional ARPES employing vacuum ultraviolet (VUV) photon energies $h\nu$ below 100 eV suffers from ill definition of surface-perpendicular momentum $k_\perp$ [6] intrinsically limited by non-free-electron final-state dispersions and large final-state broadening $\Delta k_\perp \sim \lambda^{-1}$ connected with small photoelectron escape depth $\lambda$ of a few Å. Although the previous VUV-ARPES studies [7–12] have established a general picture of the 3D-LSMO electronic structure, this limitation [7,8] as well as the use of thin film samples with their potential electronic structure distortion near the surface [13] have restricted relevance of the experimental results. Furthermore, the absence of any complete determination of the FS topology left doubts on whether the polaronic (bosonic) coupling could not disintegrate the FS in certain **k**-space regions which is one of the scenarios to explain the FS arcs in cuprates [14].

In this Letter, we explore the electronic structure of single-crystal 3D-LSMO with the optimal Sr doping $x = 0.33$ using soft-X-ray ARPES (SX-ARPES) with $h\nu$ up to 1 keV, where the increase of $\lambda$ towards ~15 Å results in sharp definition of $k_\perp$ and thus 3D momentum [15]. We achieve the most fundamental electronic structure information about 3D-LSMO, including the 3D topology of its FS and strength of electron-polaron coupling, and consolidate these findings with theoretical models. Furthermore, we identify spectroscopic signatures of the RH lattice distortion and analyze its effect on the Jahn-Teller (JT) distortion and CMR effect.

*Experiment*. – The experiments were performed at the ADRESS beamline of the Swiss Light Source. High photon flux above $10^{13}$ photon/sec/0.01% combined with the optimized geometry of the SX-ARPES endstation [16] overpowered the dramatic loss of photoexcitation cross-section at soft-X-ray energies (a factor of ~200 for the valence Mn $3d$ states compared to VUV-ARPES) [15]. The combined energy resolution was ~120 meV. The sample was kept at 11 K to avoid the destructive effects of electron-phonon scattering on the coherent ARPES signal [17]. Variations of the emission angle $\vartheta$ and $h\nu$ were rendered into **k**-space with correction for the photon momentum $p^{ph} = h\nu/c$ [16] and using an empirical inner potential $V_0$ of 7 eV. The present experimental data were collected with circular X-ray polarization.

Single crystals of La$_{1-x}$Sr$_x$MnO$_3$ with the optimal Sr doping $x = 0.33$ were grown by the floating zone method at the Moscow Power Institute. Their electric resistivity and magnetic susceptibility characterization was in agreement with the published data [18]. The crystals were FM metals below $T_C = 360$ K and poorly conducting PM metals above. Structural characterization with X-ray diffraction, see the Supplemental Material [19], has identified their RH-distorted cubic structure. The cubic-lattice equivalent lattice constant is $a = 3.89$ Å almost independent of temperature. The samples were cleaved *in situ* by an anvil-knife setup. The resulting (001) surfaces were optically rough but well defined electronically as evidenced by clear LEED images without any signs of surface reconstructions.

*Theoretical electronic structure*. – We used the standard DFT framework with the exchange–correlation functional treated within the generalized gradient approximation (GGA). The local correlation effects necessary to describe the half-metallic nature of 3D-LSMO were introduced via the empirical Hubbard parameter $U = 2$ eV applied to the Mn $3d$-states [8]. The band calculations used the full-potential APW method implemented in the Wien2k package. The Sr doping randomly replacing La atoms was introduced with the virtual crystal approximation (VCA), for details see

Ref. [8]. Fig. 1 shows the theoretical band structure $E(\mathbf{k})$ and FS calculated for the ideal cubic structure with $a = 3.89$ Å. As the RH-distortion is relatively weak, these calculations will serve as a roadmap for our further analysis. The (spin-up) valence bands near the Fermi level $E_F$ are defined essentially by the Mn $3d$ states of the $e_g$ type (almost insensitive to the variations of $U$) which split into the $3z^2 - r^2$ and $x^2 - y^2$ states and hybridize with primarily the O $2p$ states [20]. The theoretical FS consists of the sphere-like electron pocket ($e$-spheroid) around the $\Gamma$-point and cube-like hole pockets ($h$-cuboids) centered around the Brillouin zone (BZ) corners. In a perspective of our photoemission analysis, we have also performed the GGA+$U$ calculations using the fully relativistic Korringa–Kohn–Rostoker (KKR) method as implemented in the Munich SPR-KKR code [21] with the Sr doping introduced within the coherent potential approximation (CPA) that correctly reproduces the $E(\mathbf{k})$ smearing due to the disordered nature of 3D-LSMO. The APW+VCA and KKR+CPA calculations yielded indistinguishable band structures.

*Experimental band dispersions and polaronic coupling.* – The experimental results representing the valence band spectral function $A(\omega,\mathbf{k})$ along two directions in the BZ are shown in Fig. 2 as images of the ARPES intensity depending on binding energy $E_B$ and surface-parallel momentum $k_x$, (along the analyzer slit). To emphasize the dispersive spectral structures, we have subtracted from the raw ARPES intensity its non-dispersive component obtained by angle integration, for the raw data see [19]. The data along the $\Gamma X_X$ and $\Gamma M$ lines were measured with $h\nu = 643$ eV chosen at the main Mn $2p$ absorption peak to resonantly enhance photoemission intensity from the Mn $3d$ valence states. Incidentally, the corresponding $k_\perp$ values (indicated on top of the images) slightly varying with $k_x$ appear close to $8\cdot(2\pi/a)$ at the $\Gamma X_X$ and $\Gamma M$ lines. Comparison of the experimental data with the theoretical $E(\mathbf{k})$ in Fig. 1 immediately identifies the $e_g$ bands derived from the $3z^2 - r^2$ orbitals (forming the $e$-spheroid of the FS) and the $x^2 - y^2$ orbitals. In the $\Gamma X_X$ data, we note that $k_\perp$ comes closest to the exact $\Gamma X_X$ line in the second surface BZ (marked $\Gamma_1$), which results in a visually deeper $3z^2 - r^2$ band compared to the first BZ ($\Gamma_0$). In contrast to the previous VUV-ARPES data [7–11] (for a detailed analysis of the differences see Ref. [19]) our experiment shows remarkable agreement with the GGA+$U$ calculations, Fig. 1 (*a*), with only a slight $e_g$ bandwidth renormalization of around 20%.

ARPES studies of the bilayer compounds $La_{2-2x}Sr_{1+2x}MnO_7$ have found $A(\omega,\mathbf{k})$ having a pronounced peak-dip-hump structure [3] composed of the quasiparticle (QP) peak and polaronic hump. In contrast, our data in Fig. 2 – also see the energy distribution curves (EDCs) in Ref. [19] – demonstrate single peaks (albeit broadened beyond the lifetime due to the remnant $\Delta k_\perp$ effects [7,8] and intrinsic disorder of the intermixed La and Sr ions in 3D-LSMO). The peaks disperse up to $E_F$ without any loss of the spectral weight. This identifies them as the QP peaks of $A(\omega,\mathbf{k})$ because the humps, for all known polaronic systems, may disperse but always stay below $E_F$ [3]. This conclusion is confirmed by the ARPES temperature dependence [19]. The hump can only be suspected in slight asymmetry of the spectral peaks, with its vanishing weight being consistent with the insignificant band renormalization. The weak polaronic coupling in 3D-LSMO reflects its 3D nature where electron hopping to a larger number of the nearest Mn neighbors compared to the layered LSMO impedes stabilization of the polarons.

*Experimental FS*. – Figures 3(*a*) and 3(*b*) show the experimental FS maps representing the $\Gamma X_X M$ and $X_Z M'R$ horizontal planes of the BZ. The maps were measured as the ARPES intensity at $E_F$ as a function of the surface-parallel momenta $k_x$ and $k_y$ (the latter varied through the sample rotation). The $\Gamma X_X M$ map was measured again at the Mn 2*p* resonance. We immediately recognize the characteristic circular cuts of the *e*-spheroids around the $\Gamma$-points expected from the theoretical FS in Fig. 1 (*b*). The $X_Z M'R$ map measured with $hv$ varying around 708 eV immediately shows the characteristic cubic cuts of the *h*-cuboids around the R-points in Fig. 1 (*b*). We note that the remnant final-state $\Delta k_\perp$ broadening builds up, in the $\Gamma X_X M$ map, the checkerboard intensity enhancements over the M-points projected from the *h*-cuboids and, in the $X_Z M'R$ map, the enhancements over the $\Gamma$-points projected from the *e*-spheroids.

Figures 3(*c*) and 3(*d*) show the experimental FS maps measured in the vertical $\Gamma X_X M'$ and $X_Y MR$ planes under variation of $k_\perp$ through $hv$. These $k_\perp$-maps show the same characteristic pattern of the spheres and cubes as the above $\mathbf{k}_{//}$-cuts, as expected because of the essentially cubic structure of 3D-LSMO. We only note somewhat larger broadening of the horizontal FS fragments compared to the vertical ones which again manifests the $\Delta k_\perp$ effects. Therefore, the whole body of our results fully confirms the canonical FS topology in Fig. 1 (*b*). To the best of our knowledge, our experiment is the first complete resolution of this most fundamental property of the electronic structure of 3D-LSMO. The FS does not show any evident regions of suppressed intensity, which might otherwise have suggested destructive enhancements of **k**-dependent polaronic (bosonic) coupling.

*Shadow FS and its origin*. – The most intriguing discovery in the experimental FS maps in Fig. 3 are distinct replicas of the *h*-cuboids seen around the *e*-spheroids (most pronounced in the $\Gamma X_X M$ map) and, vice versa, those of the *e*-spheroids inside the *h*-cuboids. They can be represented as a translation of the fundamental FS in Fig. 1 (*b*) with an umklapp vector $(\pi/a, \pi/a, \pi/a)$ along the space diagonal $R\Gamma R$ of the BZ to halve its size. As these replicas can be viewed as a 3D analogue of the similar effect in cuprates [22,23] we will call them also "shadow" FS (SFS). In particular, the umklapp translates the RM'R direction onto the horizontal $\Gamma X_Y \Gamma$ one. Correspondingly, the $x^2 - y^2$ band around the M'-point, see Fig. 1 (*a*), produces the shadow band around the X-point, Fig. 2 (*a*), to form the SFS *h*-cuboids.

No surface reconstruction could induce this effect, because in that case the corresponding umklapp vector would have been parallel to surface, plus our samples showed no surface reconstructions. An appealing explanation of the SFS would be some hidden magnetic order potentially related to the CMR of 3D-LSMO. Such a possibility has been long debated in connection with the SFS in cuprates [22]. Another possibility would be an orbital and/or charge order [24]. However, the spectroscopic signatures of these two order parameters are normally extremely weak. The most plausible explanation is then some structural effect. We note that the long debated SFS in cuprates has finally proved to have the same origin [23]. Indeed, the structural distortion modes involving tilting of the atomic arrangement in the unit cell of the 3D-LSMO with $x = 0.33$ transform the cubic to RH-distorted lattice structure [25]. The resulting unit cell in Fig. 4 (*a*) doubles the cubic unit cell along the space diagonal, or halves the cubic BZ in the R$\Gamma$R direction as observed in our experiment. This mechanism is further corroborated by relaxation of the ARPES linear dichroism which reflects collapsing symmetries of the valence states [19].

This interpretation is confirmed by results of our relativistic one-step photoemission calculations extended to the alloy systems which included the KKR+CPA initial states, photoemission matrix elements and final states effects [26,27]. The embedded RH-distortion model included displacement of the apical oxygen atoms below and above the (001) oriented Mn-O planes (see the structural parameters in Ref. [19]). The ARPES intensity was calculated as $I_{ARPES} \propto \langle \phi^f_\mathbf{K} | \Delta^* \mathrm{Im}\, \mathbf{G}_i \Delta | \phi^f_\mathbf{K} \rangle$, where $\phi^f_\mathbf{K}$ represents the final state as the time reversal LEED state, $\Delta = A_0 \mathbf{p}$ is the dipole operator where $A_0$ denotes constant amplitude of the electromagnetic field vector potential and **p** is the momentum operator, and the Green's function $\mathbf{G}_i$ represents the initial valence states $\phi^i_\mathbf{k}$ [26]. In the simplest approximation this expression represents basically the matrix element weighted projection of $\phi^i_\mathbf{k}$ onto the final-state represented by the plane wave $\phi^f_\mathbf{K} = e^{i\mathbf{Kr}}$ as $I_{ARPES} \propto \langle e^{i\mathbf{Kr}} | \phi^i_\mathbf{k} \rangle$. Therefore, our photoemission analysis is essentially the direct plane-wave projection [28] of the rhombohedral lattice wave functions to yield the valence band $A(\omega,\mathbf{k})$.

Figures 4 (*b*) and 4(*c*) show our ARPES calculations of the FS maps in the $\Gamma X_X M$ and $X_Z M'R$ planes for the ideal cubic and RH-distorted lattice. As expected, for the cubic case these maps show

the canonic FS contours identical to the GGA+$U$ calculations in Fig. 1 (*b*). Notably, perfectly reproduced are the $\Delta k_\perp$ effects, the projections of the *h*-cuboids in the $\Gamma X_X M$ plane and those of the *e*-spheroids in the $X_Z M'R$ plane. When we turn on the RH-distortion, we immediately see the contours of the SFS *h*-cuboids and *e*-spheroids in complete agreement with the experiment in Figures 3(*a*) and 3(*b*) not only on their position but also relative intensities. Our calculations therefore confirm the RH lattice distortion as the origin of the SFS.

*Connection of the RH-structural distortion with the JT-distortion.* On first glance, the atomic displacements caused by the RH-distortion should change the atomic force constants and therefore the strength of the polaronic coupling. However, this distortion leads the manganese ion to a local symmetry of $\bar{3}$ which splits the $t_{2g}$ levels but does not remove the degeneracy of the $3z^2 - r^2$ and $x^2 - y^2$ states forming the FS. The RH-distortion does not therefore affect the JT-activity and polaronic coupling in the first order, with the higher-order corrections being less important. On the other hand, although the RH-distortion hardly changes the length of each Mn-O bond, it reduces the Mn-O-Mn angle from an ideal 180° to a significantly smaller value of 166.46°, which weakens the DE interaction between the two Mn atoms (expressed by the effective electron hopping $t_{eff}$) and thus reduces the electron itineracy. According to the theory of Millis *et al.* [1,2] the crossover from the FM metal to the poorly conducting PM metal occurs when electron itineracy is prevailed by the polaronic self-trapping. The RH-distortion has therefore an effect to shift the CMR to lower $T_c$. In this respect, the RH-distortion acts opposite to the trend when going from the layered LSMO compounds to the 3D ones where the increase of dimensionality and coordination number of the Mn-O-Mn bonds facilitates the electron hopping and thus suppresses the polaronic self-trapping.

*Conclusion.* – We have explored the electronic structure of 3D-LSMO using SX-ARPES with its intrinsically sharp definition of 3D electron momentum and single-crystal samples free of the surface reconstructions. The experimental valence band $A(\omega,\mathbf{k})$ demonstrates considerably weaker electron-polaron coupling compared to the bilayer LSMO compounds. Obscured in the previous VUV-ARPES studies, the experimental FS appears as the canonic 3D manifold of electron *e*-spheroids and *h*-cuboids predicted by GGA+$U$ band calculations. Apparent shadow FS structures, analyzed with one-step photoemission calculations, prove to manifest the RH lattice distortions rather than any magnetic or orbital order. The RH distortion is neutral to the JT-effect and thus polaronic coupling but reduces the electron itineracy and therefore acts to shift the CMR to smaller $T_C$.

*Acknowledgements.* – We thank V.A. Gavrichkov for promoting discussions, A.M. Balbashev for giving us access to single crystals, and S.S. Johnston for expert advice on polaronic physics. J.M. was supported by the Deutsche Forschungsgemeinschaft through Grant FOR 1346, the BMBF (Project 05K13WMA) and CENTEM (CZ.1.05/2.1.00/03.0088). J.M. and P.B. thank European COST network Euspec (MP1306) for support. A.S.M. acknowledges the support of ImPACT Program of Council for Science, Technology and Innovation (Cabinet Office, Government of Japan).

———————————————

[*] lll_ru@mail.ru

[†] vladimir.strocov@psi.ch

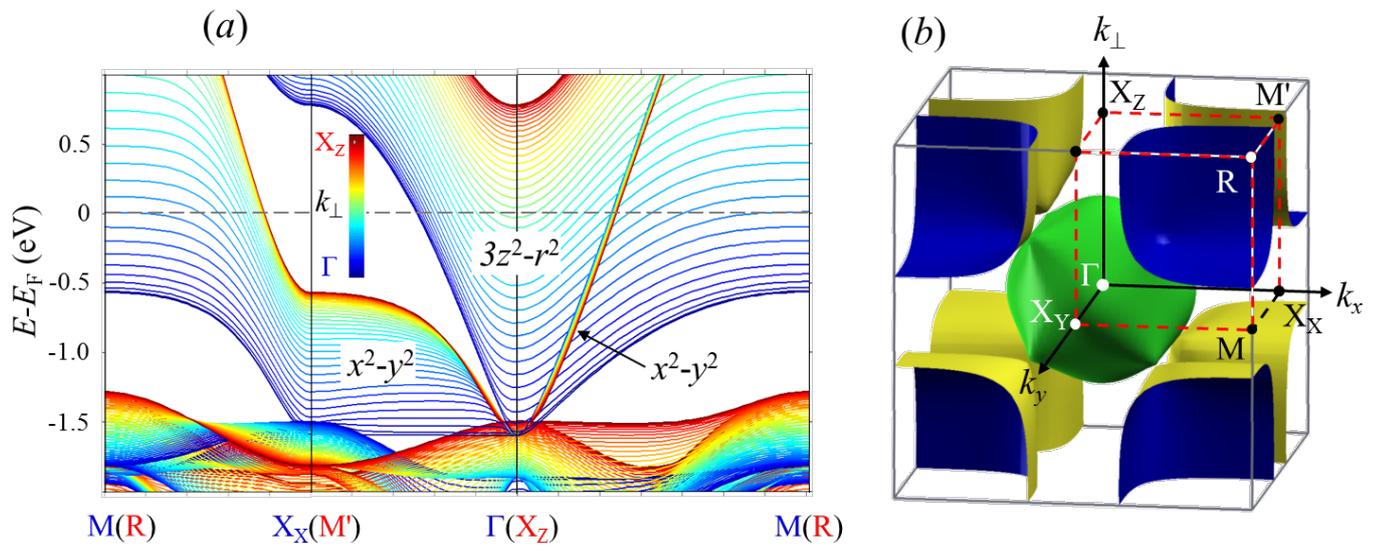

Fig. 1. GGA+$U$ calculations for the ideal cubic structure of 3D-LSMO [7,8]: (*a*) (Spin-up) band structure, with the color scale indicating $k_\perp$ of the bands; (*b*) Fermi surface.

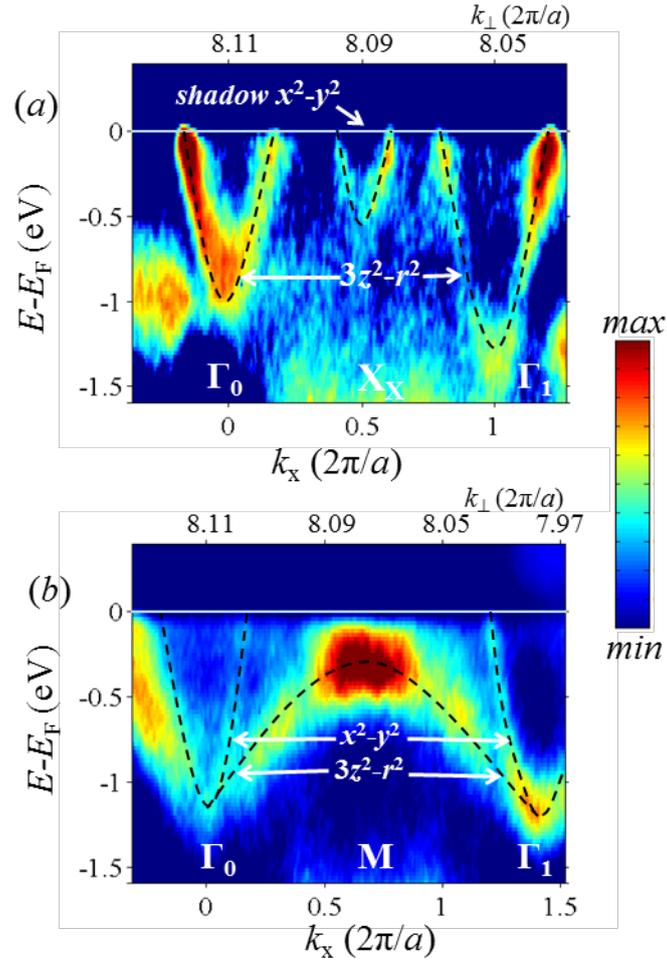

Fig. 2 (color online). Experimental ARPES images representing the $\Gamma X_X$ (*a*) and $\Gamma M$ (*b*) lines of the BZ measured with $h\nu = 643$ eV (Mn 2*p* resonance). The $k_\perp$ variations are shown on top of the images. The dashed lines are eye guides for the experimental dispersions. The orbital character of the bands is indicated, and the shadow $x^2 - y^2$ band is marked. The experimental bands do not show any significant renormalization compared to the GGA+$U$ bands in Fig. 1 (*a*).

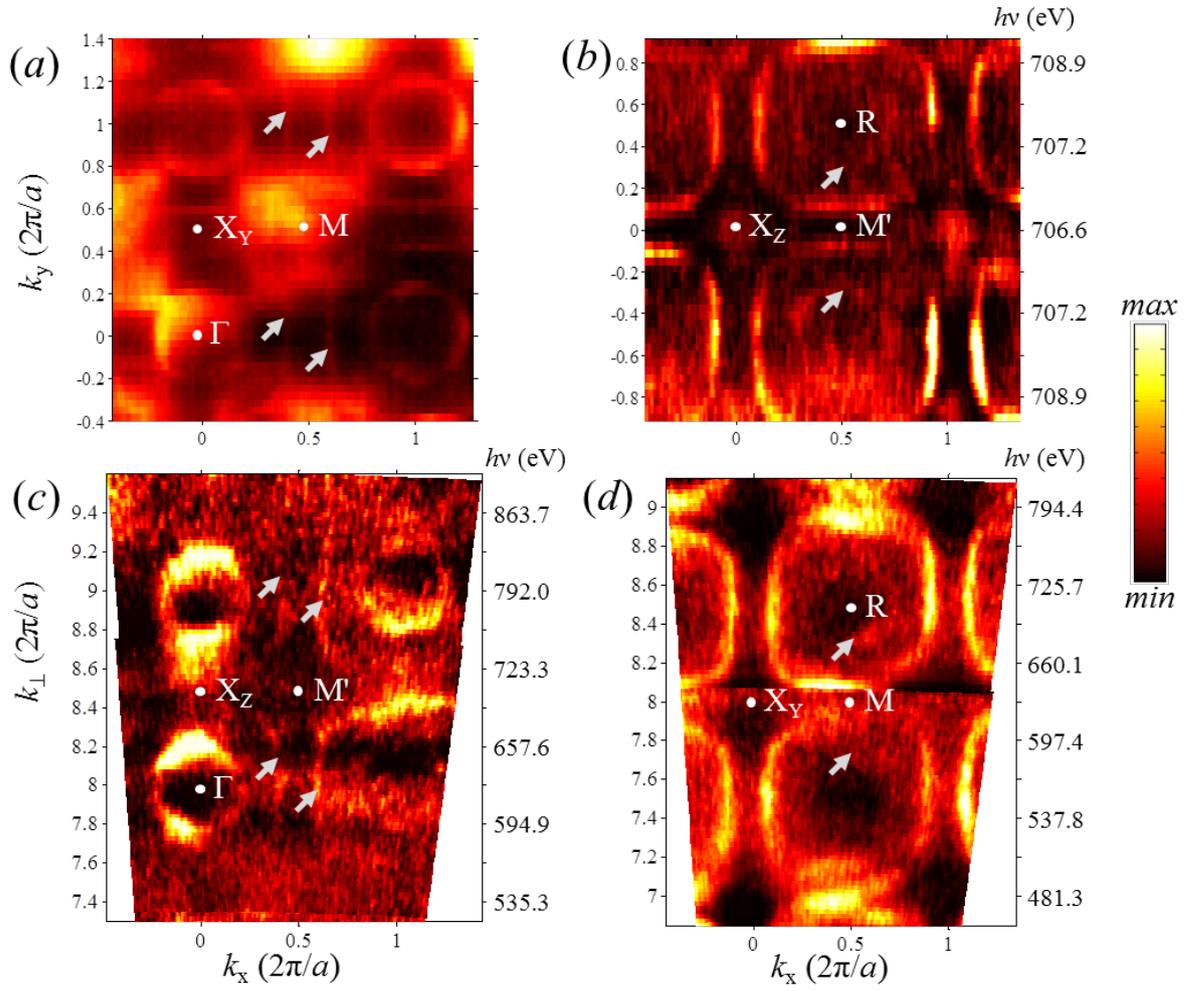

Fig. 3 (color online). Experimental $k_{//}$-cuts of the FS measured as a function of $\theta$, representing the $\Gamma X_X M$ plane (*a*, measured at $h\nu = 643$ eV) and $X_Z M'R$ plane (*b*, with $h\nu$ marked on the right varying around 708 eV as a function of $k_y$ to keep constant $k_\perp$); experimental $k_\perp$-cuts measured under $h\nu$ variations (marked on the right for $k_x = 0$) representing the $\Gamma X_X M'$ (*c*) and $X_Y MR$ (*d*) planes. For the latter, $h\nu$ tracked the sample rotation to keep constant $k_y = \pi/a$. The experiment fully confirms the 3D topology of the FS in Fig 1 (*b*). The shadow FS contours are marked by arrows.

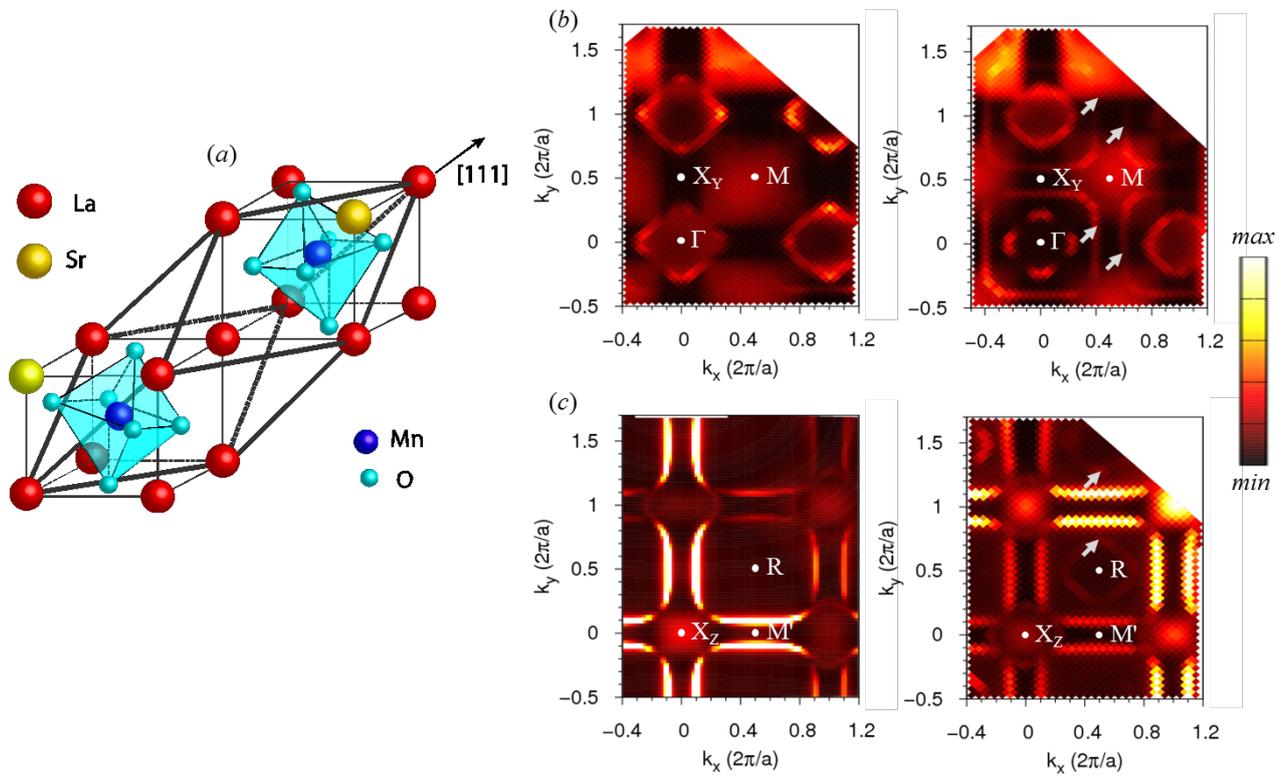

Fig. 4 (color online). (*a*) RH-distortion of the 3D-LSMO unit cell with Sr randomly substituting La ions; (*b,c*) ARPES calculations of the $\mathbf{k}_{//}$-cuts representing the FS in the $\Gamma X_X M$ (*b*) and $X_Z M'R$ (*c*) planes for the ideal cubic (*left*) and RH distorted crystal lattice (*right*). The calculations confirm the structural origin of the SFS marked by arrows.

# Supplemental Material for "Fermi surface of three-dimensional La$_{1-x}$Sr$_x$MnO$_3$ explored by soft-X-ray ARPES: Rhombohedral lattice distortion and its effect on magnetoresistance"

L. L. Lev,[1,2,*] J. Krempasky,[1] U. Staub,[1] V.A. Rogalev,[1] T. Schmitt,[1] M. Shi,[1] P. Blaha,[3] A. S. Mishchenko,[4,2] A. A. Veligzhanin,[2] Y. V. Zubavichus,[2] H. Volfová,[5] J. Braun,[5] J. Minár,[5,6] and V.N. Strocov1,†

[1] Swiss Light Source, Paul Scherrer Institute, 5232 Villigen, Switzerland

[2] National Research Centre „Kurchatov Institute", 123182 Moscow, Russia

[3] Institut für Materialchemie, Technische Universität Wien, A-1060 Wien, Austria

[4] RIKEN Center for Emergent Matter Science, 2-1 Hirosawa, Wako, Saitama 351-0198, Japan

[5] Department Chemie, Ludwig-Maximilians-Universität München, 81377 Munich, Germany

[6] New Technologies-Research Center, University of West Bohemia, Plzen, Czech Republic

*X-ray diffraction characterization*. – X-ray diffraction measurements of our 3D-LSMO samples with Sr concentration $x = 0.33$ have been performed at the Structural Materials Science beamline [1] of the Kurchatov Synchrotron Radiation Source, Moscow. The measurements used the Fujiifilm Imaging Plate detector. The X-ray wavelength was set to 0.68886 Å. For room-temperature measurements, the 2D diffraction patterns were acquired within an angular range from 3 to 50 degrees. For low-temperature measurements, the sample was placed on the cold head of the Simutomo Heavy Industries SDRK-408D cryocooler. In this case the diffracted intensity was measured through a narrow window in the vacuum chamber, allowing the registered angular range from 5 to 30 degrees. The silicon (SRM 640e) and corundum (SRM 676a) NIST standards were used for calibration of the diffraction range. The diffraction patterns were integrated into 1D diffraction curves with the Fit2D software [2].

The X-ray diffraction curves measured on powder LSMO samples at $T = 300K$ and 4.7K are shown in Fig. SM1. Large intensity changes between the two curves are caused by preferential orientations in the powder. The experimental curves are compared with the one simulated for the RH distorted crystal lattice adopting the R$\bar{3}$c symmetry group. The simulations used the Crystallographica software [3]. The reflections splittings in the indicated regions around 17.5, 23 and 29° seen for both temperatures are characteristic of the R$\bar{3}$c group. Therefore, our X-ray diffraction results clearly identify the RH lattice distortion as an intrinsic crystal structure property of the 3D-LSMO crystals in the region of Sr concentrations around $x = 0.33$. The splittings increase with decrease of temperature, which suggests corresponding strengthening of the RH distortion. The cubic-lattice equivalent lattice constant has been measured as 3.89 Å at $T = 300$ K. It shows only an insignificant decrease around 0.2% towards low temperatures [4].

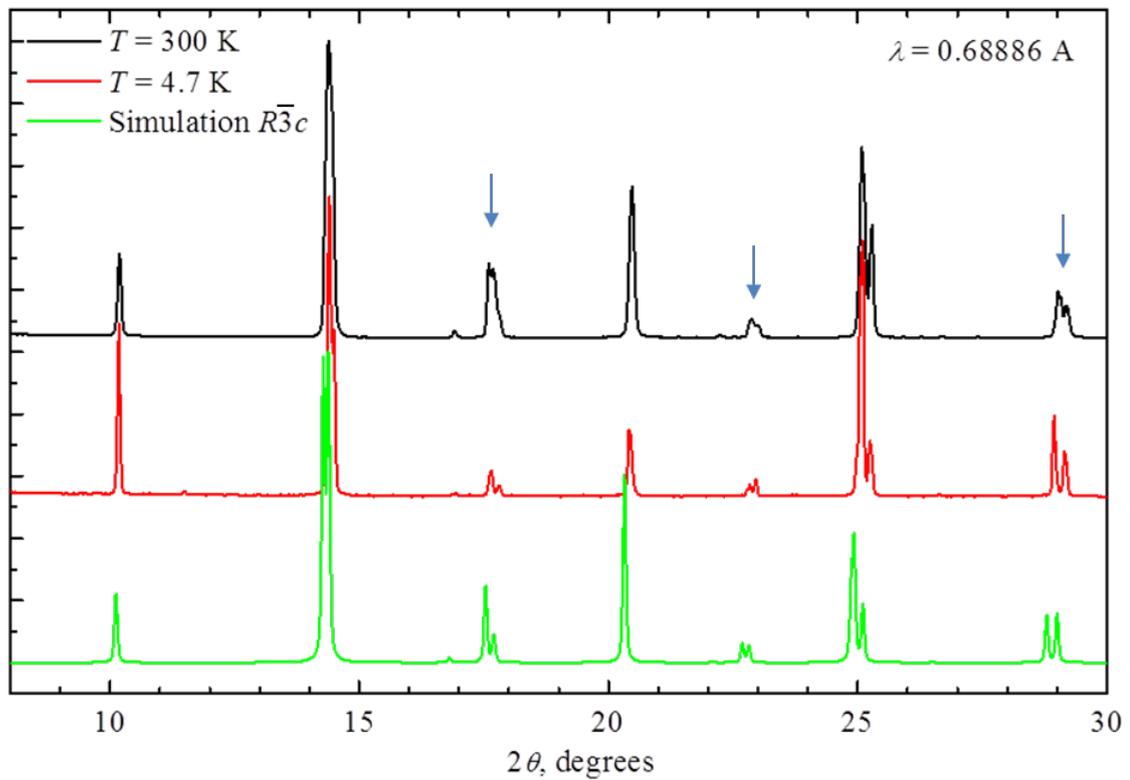

Fig. SM1. Experimental 1D diffraction curves of the 3D-LSMO samples measured at two temperatures compared with simulations for the $R\bar{3}c$ symmetry group. The arrows mark the regions identifying the RH distortion.

*Raw experimental data.* – The EDCs corresponding to the ARPES images in Fig. 2 are shown in Fig. SM2. (*a*) and (*c*) show the raw EDCs (without smoothing) for the $\Gamma X_X$ and $\Gamma M$ lines of the BZ, respectively. The common visual aspect of these EDCs, in particular for the $\Gamma X_X$ line, is large non-dispersive intensity which originates from the Mn $t_{2g}$ states and scales up towards high $E_b$. To emphasize the dispersive spectral structures originating from the Mn $e_g$ states, we have subtracted from the raw EDCs a non-dispersive component obtained by their integration over the whole shown angular range. Representing the dispersive component of the ARPES intensity, these EDCs shown in (*b*) and (*d*) were used to generate the ARPES images shown in Fig. 2.

We note that the previous VUV-ARPES studies on thin films of 3D-LSMO [5–11] were systematically finding broad band dispersions and allegedly strongly renormalized band dispersions, with the intensity decay towards $E_F$ precluding accurate resolution of the FS. Later works [5,6] suggested that these effects could result from ill 3D momentum definition of low-energy ARPES final states, with additional complications due to surface reconstructions and inherent electronic structure distortions of the thin film 3D-LSMO [12]. This is confirmed by the sharp spectral structure yielded by our SX-ARPES experiment in agreement with the GGA+$U$ calculations.

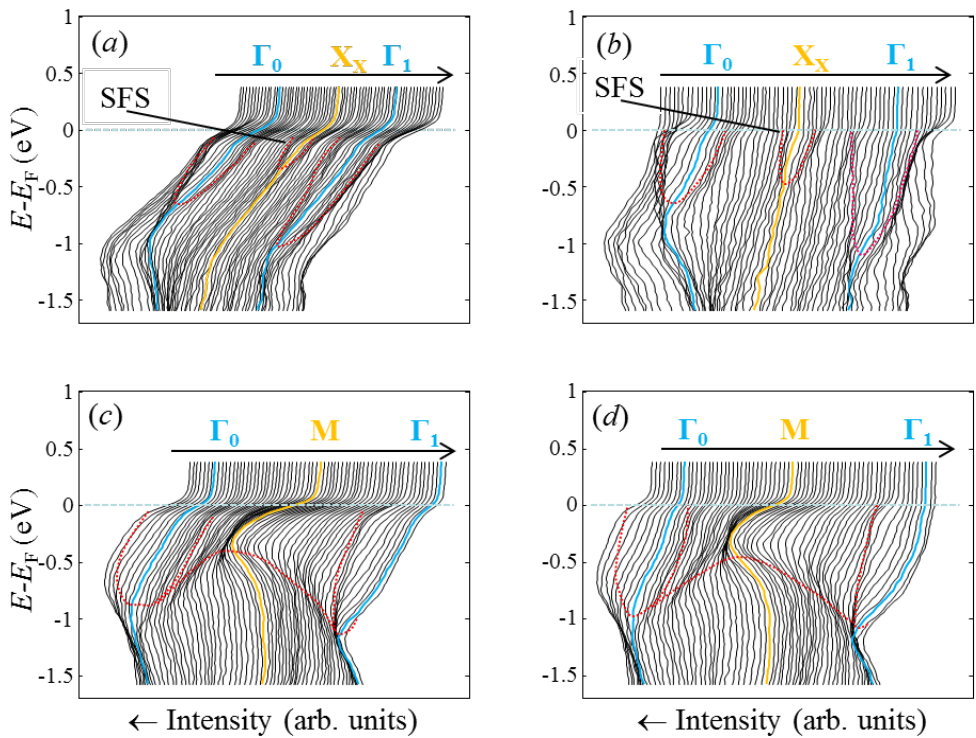

Fig. SM2. The raw EDCs corresponding to the ARPES images in Fig. 2 (*a*, *c*) and their angle-dispersive component (*b*, *d*). The red lines indicate the $x^2 - y^2$ and $3z^2 - r^2$ bands marked in Fig. 2. The structures corresponding to the shadow $x^2 - y^2$ band forming cuboid are marked as SFS (*a*, *b*).

*Temperature dependence and polaronic satellites.* – Typical spectral function $A(\omega,\mathbf{k})$ in the polaronic materials like LSMO is characterized by a pronounced peak-dip-hump structure where the narrow peak corresponds to the QP excitation and broader hump to its polaronic tail [13,14]. Here, we endeavor temperature dependent measurements to elucidate the QP vs polaronic nature of the experimental $A(\omega,\mathbf{k})$ peaks observed by ARPES for our 3D-LSMO compound. We will analyze these data in comparison with the bilayer $La_{2-2x}Sr_{1+2x}MnO_7$ compound extensively studied by a variety of techniques including ARPES [13] as well as X-ray core level photoemission (XPS), absorption (XAS) and emission (XES) spectroscopies [15]. For that compound, the ratio between the QP peak and polaronic hump in $A(\omega,\mathbf{k})$ reduces with temperature [13] which reflects the competition between the DE related electron itineracy and polaronic self-trapping [16,17].

For analysis of the temperature dependent $A(\omega,\mathbf{k})$ in 3D-LSMO, we have chosen the $x^2 - y^2$ band along the $X_ZR$ symmetry line of the BZ, see Fig. 1 (*a*). This band has purely 2D character, which excludes extrinsic broadening of its ARPES response owing to the final-state $\Delta k_\perp$ broadening [18], and is well separated from the $3z^2 - r^2$ continuum. Under these conditions the ARPES peaks can be reconciled with the true valence band $A(\omega,\mathbf{k})$. For intensity reasons, the measurements used *s*-polarization of the incident X-rays, with $hv = 643$ eV chosen at the Mn 2*p* resonance. The $X_ZR$ line was then approached by rotation of the sample towards the $\Gamma_{22}$ point in the surface reciprocal lattice, which delivered $k_z = 7.57 \cdot (2\pi/a)$ close to this line.

The resulting ARPES data at $T = 12$K are shown in Fig. SM3, represented as the intensity map (*a*) as well as the EDCs (*c*, *black line*) drawn as indicated in (*a*) around the middle of the $x^2 - y^2$ bandwidth. The EDCs are averaged over the four equivalent cuts and normalized to integral intensity within the shown $E_B$ range. In stark contrast to the bilayer LSMO, the observed $A(\omega,\mathbf{k})$ is characterized by a single albeit rather broad peak dispersing up to $E_F$ without any loss of the spectral weight. We can only guess a touch of its asymmetry towards lower $E_b$ but such small effect can be confused with the colossal spectral intensity protruding from the $t_{2g}$ states below the $e_g$ ones.

To distinguish between the QP peak vs polaronic bump nature of the observed $A(\omega,\mathbf{k})$, we gradually increased temperature to $T = 300$K. Fig. SM2 (*b*) shows the map at 300K, and (*c*) the whole series of the EDCs throughout the temperature ramp. Even within this colossal interval, the observed evolution of $A(\omega,\mathbf{k})$ is much weaker compared to the bilayer LSMO [13], with its notable asymmetry growing on the low-energy side rather attributed to the protruding $t_{2g}$ intensity much increasing with temperature. If the $A(\omega,\mathbf{k})$ peak had been the polaronic hump, quite the opposite, its broadening must have dramatically increased owing to broadening of the phonon spectrum [13]. Therefore, the observed temperature dependence suggests that the experimental $A(\omega,\mathbf{k})$ reflects essentially the QP peak. We note however that its broadening is much above the usual electron lifetime effects and has rather a Gaussian but not Lorentzian shape. This can be explained, at this moment only tentatively, by creation of polarons losing their coherence due to scattering in the inhomogeneous 3D media of our crystal where La and Sr ions are intermixed. In this case the $A(\omega,\mathbf{k})$ broadening should decrease with temperature because of the motional narrowing [19] but this effect is presumably counteracted by extrinsic spectral broadening due to the thermal photoelectron scattering at our high photoelectron energies [20].

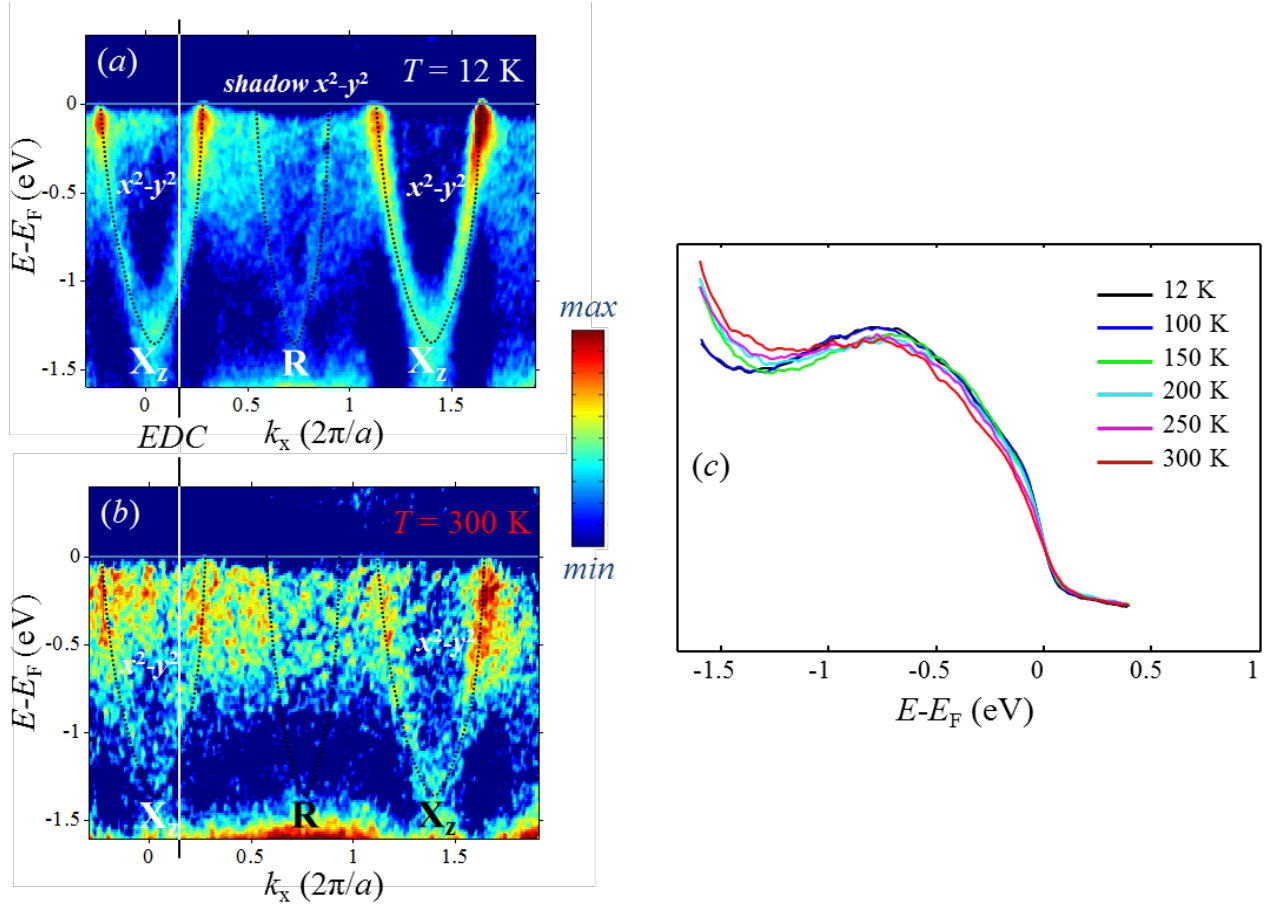

Fig. SM3. (a,b) ARPES images along the $X_ZR$ line of the BZ measured with $h\nu = 643$ eV at $T = 12$K and 300K, respectively; (c) Series of EDCs along the cut indicated in (a,b) throughout the temperature ramp. The $x^2 - y^2$ spectral peak does not much evolve with temperature, which identifies its QP origin.

Only vague traces of the polaronic tail of the QP peak as well as insignificant renormalization of the ARPES dispersions compared to the GGA+$U$ bands, Fig. 1 (a), indicate weaker polaronic coupling in our 3D-LSMO compound compared to the layered LSMO. This is directly related to its 3D structure where increase of the electron hopping perpendicular to the layers as expressed by the $z$ parameter impedes stabilization of the polarons as compared to the layered LSMO.

The absence of any clear spectroscopic signatures of polarons in our ARPES data is seemingly in conflict with those of mid-infrared optical spectroscopy, pulsed neutron scattering and extended X-ray absorption fine structure (EXAFS) experiments on 3D-LSMO [17,21,22] which observed weak to moderate polaronic effects existing already in the low-temperature region of the phase diagram and strengthening with increase of temperature. However, none of these techniques is sensitive to inherent inhomogeneity of the 3D-LSMO system [23] which is proved by experiments sensitive to the local lattice distortions [21,22]. On the contrary, the momentum selectivity of ARPES makes it dramatically sensitive to the inhomogeneity, hence the polaronic shape distortion can easily be overshadowed by the inhomogeneity induced spectral broadening.

The present work has been focused on the 3D-LSMO compound with the optimal Sr concentration $x = 0.33$ delivering maximal $T_c$. An extension of our experimental methodology can be envisaged to explore the rich interplay of the charge, orbital, spin and lattice degrees of freedoms throughout the whole phase diagram of manganites. We conjecture that in 3D-LSMO compounds with lower Sr concentrations the decrease of $T_c$ can be caused by strengthening of the polaronic effects despite the three-dimensionality. In this case the polaronic effects may already be observed in ARPES as the characteristic peak-dip-hump structure. In particular, we envisage an extension of our experiments to the concentration $x = 0.175$ whose $T_c$ is reduced to 260 K and the resistance above $T_c$ increases by about two orders of magnitude compared to our $x = 0.33$. Furthermore, these crystals undergo the orthorhombic lattice distortions as opposed to the RH one in our case. For these reasons such 3D-LSMO system should exhibit an interplay of the polaronic and lattice distortion effects different from our present compound.

*Comparison of the SX- and VUV-ARPES data.* – Our results, measured with SX-ARPES on single-crystal LSMO samples, are significantly different from the previous VUV-ARPES results on thin-film samples [5–11]. Fig. SM4 compares our experimental dispersions along the $\Gamma X_X$ direction with the VUV-ARPES ones from Ref. [9]. In the $\Gamma_1$-point, bringing $k_\perp$ into the exact $\Gamma X_X M$ plane, our $3z^2 - r^2$ bandwidth appears about 0.5 eV deeper. This brings our results in remarkable agreement with the GGA+$U$ band structure in Fig. 1 with only an insignificant band renormalization of around 20% as compared to around 50% claimed in Ref. [7,8].

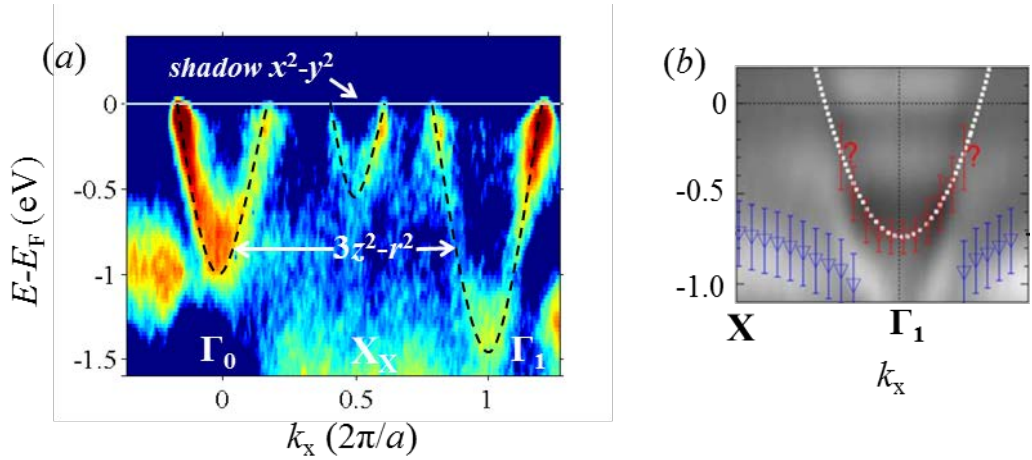

Fig. SM4. Comparison of the experimental SX-ARPES (*a*) and VUV-ARPES (*b*) [9] dispersions along the $\Gamma X_X$ direction. Ill $k_\perp$ definition of the low-energy final states and electronic structure distortions in strained thin films reduce the apparent VUV-ARPES bandwidth.

The differences between the two experiments are explained by (all or some of) the following factors:

(1) The VUV-ARPES bandwidth reduction is mostly an extrinsic effect caused by large $\Delta k_\perp$ broadening of the low-energy final states (for details the $\Delta k_\perp$ effects see [18]). Model VUV-ARPES calculations taking into account $\Delta k_\perp$ [5] have indeed found an upward shift about 0.3 eV in the bottom of the $3z^2 - r^2$ band. In the SX-ARPES energy range $\Delta k_\perp$ reduces by a factor of 3-5, almost eliminating this extrinsic effect;

(2) $k_\perp$ dispersions of the low-energy final states can in fact deviate from the free-electron-like model. Indeed, the experimental ARPES dispersions as a function of $hv$ [10] appear far from regular oscillations expected from free-electron-like final states [24]. The use of this model can shift $k_\perp$ from the exact $\Gamma X_X M$ plane and shift the observed $3z^2 - r^2$ band dispersion upwards;

(3) All previous VUV-ARPES works used thin-film samples of LSMO. However, Pesquera *et al.* [12] have shown that the tensile strain in thin films of manganites distorts their electronic structure near the surface, modulating the Mn $e_g$ orbital filling ultimately towards reversal of the $e_g$ band structure. In addition, the thin-film samples typically suffer from surface reconstructions which can also significantly distort their ARPES response. These problems are eliminated in the present work because of the use of single-crystal samples. We note that surfaces of the fractured LSMO samples are typically rough and show domains with dimensions below 100 um, which makes the ARPES experiment extremely challenging.

*Relaxation of the ARPES linear dichroism.* – ARPES experiments on single crystals can use the dipole selection rules with linearly polarized light – linear dichroism – to identify symmetries of the valence states. Indeed, if the incoming X-rays are *p*-polarized (parallel **E**-vector) relative to certain symmetry plane of the crystal, they will excite from this plane only the symmetric states, and if *s*-polarized (perpendicular **E**-vector), they will excite only the antisymmetric states. Tailored to our experimental geometry, these selection rules are discussed Ref. [25].

The RH-distortion along the spatial diagonal of the cubic unit cell leaves only one of its symmetry planes. Furthermore, owing to the absence of any preferential direction, the bulk crystal forms domains with three different orientations of the elongated diagonal. Therefore, macroscopically, the RH-distorted crystal has no exact symmetry planes. This immediately results in relaxation of the strict ARPES linear dichroism. Indeed, the experimental spectra in Fig. SM5 demonstrate that the $3z^2 - r^2$ and $x^2 - y^2$ bands along $\Gamma X_X$, $\Gamma M$ and $X_Z R$ directions do show intensity changes upon switching the polarization but no strictly on/off dichroism characteristic of the single crystals [25].

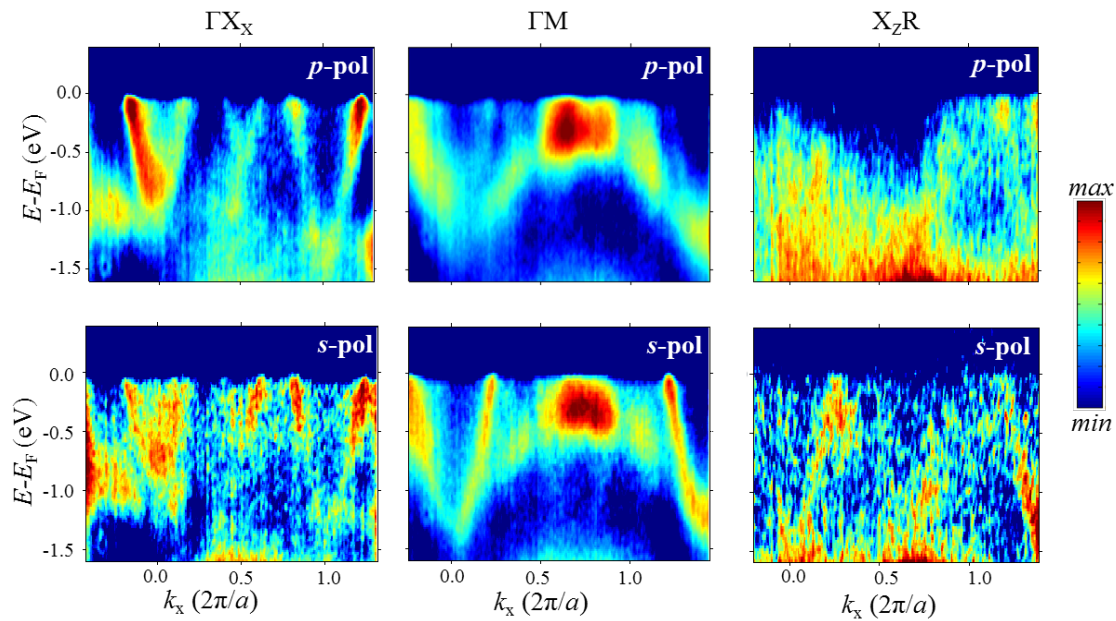

Fig. SM5. ARPES images corresponding to the $\Gamma X_X$, $\Gamma M$ and $X_Z R$ directions of the cubic cell measured with photon energy $h\nu$ = 643 eV (Mn 2*p* resonance) at *p*- and *s*-polarizations of incident X-rays (*top and bottom*, respectively). The relaxation of the linear dichroism reflects the RH lattice distortion.

We note that the dichroism relaxation has also been noted in one of the previous VUV-ARPES studies [26] but interpreted in terms of alleged predominant wavefunction symmetries in the ideal crystal structure. However, this approach is irrelevant because in the ideal crystal the wavefunctions can only be strictly symmetric or strictly antisymmetric relative to the symmetry planes.

*Structural parameters of the RH distortion.* – Table SM1 compiles the structural parameters of the RH-distorted lattice of 3D-LSMO used in the ARPES calculations. The space group is $R\bar{3}c$, and the lattice constants are $a$ = 5.498 Å and $c$ = 13.305 Å [27].

|       | x      | y | z    | Number | Occupancy      |
|-------|--------|---|------|--------|----------------|
| La,Sr | 0      | 0 | 0.25 | 6      | La 0.7, Sr 0.3 |
| Mn    | 0      | 0 | 0    | 6      | 1              |
| O     | 0.4582 | 0 | 0.25 | 18     | 1              |

___________________________

[*] lll_ru@mail.ru

[†] vladimir.strocov@psi.ch